\documentclass[conference]{IEEEtran}
\IEEEoverridecommandlockouts
\usepackage{cite}
\usepackage{mathtools}
\usepackage{amsmath,amssymb,amsfonts}
\usepackage[linesnumbered,ruled,vlined]{algorithm2e}
\usepackage{graphicx}
\usepackage{subcaption}
\graphicspath{ {./images/}}
\usepackage{textcomp}
\usepackage{xcolor}
\usepackage{url}

\DeclareMathAlphabet\mathbfcal{OMS}{cmsy}{b}{n}
\def\BibTeX{{\rm B\kern-.05em{\sc i\kern-.025em b}\kern-.08em
		T\kern-.1667em\lower.7ex\hbox{E}\kern-.125emX}}

\DeclarePairedDelimiter\floor{\lfloor}{\rfloor}

\begin{document}
	
	\title{Accelerated Stochastic Gradient for Nonnegative Tensor Completion and Parallel Implementation\\
		\thanks{All authors were partially supported by the European Regional Development Fund of the European Union and Greek national funds through the Operational Program Competitiveness, Entrepreneurship, and Innovation, under the call RESEARCH - CREATE - INNOVATE (project code : ${\rm T1E\Delta K-03360}$). }
	}
	
	\author{\IEEEauthorblockN{Ioanna Siaminou}
		\IEEEauthorblockA{\textit{School of Electrical and Computer Engineering} \\
			\textit{Technical University of Crete}\\
			Chania, Greece \\
			isiaminou@isc.tuc.gr}
		\and
		\IEEEauthorblockN{Ioannis Marios Papagiannakos}
		\IEEEauthorblockA{\textit{School of Electrical and Computer  Engineering} \\
			\textit{Technical University of Crete}\\
			Chania, Greece \\
			ipapagiannakos@isc.tuc.gr}
		\and
		\IEEEauthorblockN{\hspace{5em}Christos Kolomvakis}
		\IEEEauthorblockA{\hspace{6em}\textit{School of Electrical and Computer  Engineering} \\
			\textit{\hspace{6em}Technical University of Crete}\\
			\hspace{5em}Chania, Greece \\
			\hspace{6em}ckolomvakis@isc.tuc.gr}
		\and
		\IEEEauthorblockN{\hspace{5em} Athanasios P. Liavas}
		\IEEEauthorblockA{\hspace{5em}\textit{School of Electrical and Computer  Engineering} \\
			\textit{\hspace{5em}Technical University of Crete}\\
			\hspace{6em}Chania, Greece \\
			\hspace{6em}liavas@telecom.tuc.gr}
	}
	
	\maketitle
	
	\begin{abstract}
		We consider the problem of nonnegative tensor
		completion. We adopt the alternating
		optimization framework and solve each nonnegative matrix
		completion problem via a stochastic variation of the accelerated gradient algorithm. 
		We experimentally test the effectiveness and the efficiency of
		our algorithm using both real-world and synthetic data. 
		We develop a shared-memory implementation of our
		algorithm using the multi-threaded API OpenMP, which attains
		significant speedup. We believe that our approach is a very
		competitive candidate for the solution of very large
		nonnegative tensor completion problems.
	\end{abstract}
	
	
	\section{Introduction}
	\label{section_introduction}
	
	Tensors have recently gained great popularity
	due to their ability to model multiway data dependencies
	\cite{Kroonenberg_2008},
	\cite{Cichocki_et_al_2009}, \cite{Kolda_Bader_2009}, \cite{Sidiropoulos_et_al_2017}.
	Tensor decomposition (TD) into latent factors is very important
	for numerous tasks, such as feature selection, dimensionality reduction,
	compression, data visualization and interpretation. 
	The Canonical Polyadic Decomposition (CPD) is one of the most important tensor decomposition models.
	Tensor Completion (TC) arises in many modern applications such as machine learning, signal processing, and scientific computing. 
	
	We focus on the CPD model  and consider the nonnegative tensor completion (NTC) problem,  using as quality metric the Frobenius norm of the difference between the true and the estimated tensor. We adopt the Alternating Optimization (AO) framework,
	that is, we work in a circular manner and update each factor by keeping all other factors fixed. 
	We update each factor by solving  a nonnegative matrix completion (NMC) problem via a stochastic variant of  the accelerated (Nesterov-type) gradient \cite{Nesterov_2004}.
	
	Recent tensor applications, such as social network analysis, recommendation systems, and targeted advertising, need to handle very large sparse tensors. 
	We propose a shared-memory implementation of our algorithm that attains significant speedup and can efficiently handle very large problems. 
	
	\subsection{Related Work}
	
	Most of the papers that consider sparse TD and TC focus on unconstrained problems.  One of the earliest works is Gigatensor \cite{Kang_Papalexakis_et_al_2012}, which was followed by DFacTo \cite{Choi_Vishwanathan_2014}. In \cite{Karlsson_et_al_2015}, two parallel algorithms for the unconstrained TC have been developed
	and results concerning the speedup attained by their MPI implementations
	on a linear processor array have been reported. In \cite{Smith_Karypis_2015}, the authors introduce fine- and medium-grained partitionings for the TC problem, while \cite{Kaya_et_al_2018} 
	incorporates dimension trees into the developed  
	parallel algorithms. In \cite{Huang_Sidiropoulos_Liavas_2015} and  \cite{smith_karypis_AO_ADMM2015}, the AO-ADMM framework has been adopted for constrained matrix/tensor factorization and completion.
	In \cite{Liavas_et_al_2018}, the medium-grained approach  of \cite{Smith_Karypis_2015} was used for the solution of the nonnegative TD problem in distributed memory systems. 
	The same problem has been considered in \cite{Ballard_et_al_2018}, where the authors incorporate the dimension trees and observe performance gains due to reduced computational load. 
	The works in \cite{Blanco_et_al_2018}, \cite{Ge_et_al_2018}, and \cite{Shin_Kang_2014} make use of either the Map-Reduce programming model or the Spark engine. In \cite{Karsavuran_et_al_2021}, a hypergraph model for general medium--grain partitioning has been presented.
	
	Works that employ Stochastic Gradient Descent (SGD) on shared memory and distributed systems for sparse tensor factorization and completion include \cite{Papastergiou_Megalooikonomou_2017}, \cite{Smith_et_al_2016}, \cite{Xie_et_al_2020}, \cite{Devine_Ballard_2020}. In \cite{Papastergiou_Megalooikonomou_2017}, the authors describe a TC approach which uses the CPD model and employs a proximal SGD algorithm, that can be implemented in a distributed environment. In \cite{Smith_et_al_2016}, the authors examine three popular optimization algorithms: alternating least squares (ALS), SGD, and coordinate descent (CCD++), implemented on shared- and distributed-memory systems. They conclude that SGD is most competitive in a serial environment, ALS is recommended for shared-memory systems, and both ALS and CCD++ are competitive on distributed systems. In \cite{Xie_et_al_2020}, the authors propose a GPU-accelerated parallel TC scheme (GPU-TC) for accurate and fast recovery of missing data via SGD. Finally,  \cite{Devine_Ballard_2020} presents GentenMPl, a toolkit for sparse CPD that is designed to run effectively on distributed-memory high-performance computers. The authors use the Trilinos libraries and they present implementations of the CPD-ALS and an SGD method. 
		
	\subsection{Notation}
	Vectors, matrices, and tensors are denoted by small, capital, and calligraphic capital letters, respectively;
	for example, ${\bf x}$, ${\bf X}$, and $\mathbfcal{X}$. $\mathbb{R}^{I_1 \times \cdots \times I_N}_+$ denotes the set of $(I_1 \times\cdots \times I_N)$ nonnegative tensors. The elements of tensor $\mathbfcal{X}$ are denoted as $\mathbfcal{X}(i_1,\ldots,i_N)$.
	In many cases, we use Matlab-like notation, for example, ${\bf A}(j,:)$ denotes the $j$-th row
	of matrix ${\bf A}$. The outer product of vectors ${\bf a}$ and ${\bf b}$ is defined as
	${\bf a} \circ {\bf b}$. The Kronecker, Khatri-Rao, and Hadamard product of
	matrices ${\bf A}$ and ${\bf B}$, of compatible dimensions, are defined, respectively, as
	${\bf A}\otimes{\bf B}$, ${\bf A} \odot {\bf B}$ and ${\bf A}\circledast {\bf B}$; 
	extensions to the cases with more than two arguments are obvious. ${\bf I}_P$ denotes the $(P\times P)$ identity matrix, 
	$\|\cdot \|_F$ denotes the Frobenius norm of the matrix or tensor argument, and $({\bf X})_+$ denotes the matrix 
	derived after the projection of the elements of ${\bf X}$ onto $\mathbb{R}_+$.

\section{Nonnegative tensor completion}
	
Let $\mathbfcal{X}^o\in\mathbb{R}_+^{I_1  \times \dots \times I_N}$ be an $N$-th order tensor which admits the 
rank-$R$ CPD  \cite{Kolda_Bader_2009}, \cite{Sidiropoulos_et_al_2017}
\begin{equation}
\mathbfcal{X}^o  = \mbox{\textlbrackdbl} \mathbf{U}^{o(1)}, \dots, \mathbf{U}^{o(N)}
\mbox{\textrbrackdbl} =
\sum_{r=1}^R \mathbf{u}_r^{o(1)} \circ \dots \circ \mathbf{u}_r^{o(N)},
\label{Xo_densef}
\end{equation}
where $\mathbf{U}^{o(i)}=[\mathbf{u}_1^{o(i)} ~\cdots ~ \mathbf{u}_R^{o(i)}]\in\mathbb{R}_+^{I_i \times R}$, for $i=1,\ldots,N$.
We observe $\mathbfcal{X} = \mathbfcal{X}^o + \mathbfcal{E}$,
where $\mathbfcal{E}$ is additive noise. Let $\Omega \subseteq \{1, \ldots, I_1 \} \times \dots \times \{1, \ldots, I_N \}$ be the set of indices of the observed entries of ${\mathbfcal X}$. 
Also, let ${\mathbfcal M}$ be a tensor with the same size as ${\mathbfcal X}$, with elements 
${\mathbfcal M}(i_1,i_2, \dots ,i_N)$ equal to one or zero based on the availability of the corresponding element of ${\mathbfcal X}$. That is 
\begin{equation}
{\mathbfcal M}(i_1,i_2, \dots ,i_N) = \left\{ \begin{array}{ll}
1, & \mbox{if}~(i_1,i_2, \dots ,i_N) \in \Omega, \\
0, & \mbox{otherwise.} \end{array}
\right.
\label{M_def}
\end{equation}
We consider the NTC problem
\begin{equation}
\underset{\left\{{\bf U}^{(i)}\in\mathbb{R}^{I_i\times R}_+\right\}_{i=1}^N} \min f_{\Omega}\left(\mathbf{U}^{(1)}, \dots, \mathbf{U}^{(N)}\right)  + \frac{\lambda}{2} \sum_{i=1}^{N}\left\| \mathbf{U}^{(i)} \right\|_F^2,
\label{Problem_TC}
\end{equation}
where
\begin{equation*}
f_{\Omega}\hspace{-.07cm}\left(\mathbf{U}^{(1)}, \dots, \mathbf{U}^{(N)}\right) 
\hspace{-.05cm}=\hspace{-.05cm}\frac{1}{2} \left\| \mathbfcal{M} \hspace{-.05cm}\circledast 
\hspace{-.05cm}
\left( \hspace{-.05cm}\mathbfcal{X} \hspace{-.05cm}- \hspace{-.05cm}\mbox{\textlbrackdbl} \mathbf{U}^{(1)}, \dots, \mathbf{U}^{(N)} \mbox{\textrbrackdbl} \right) \right\|_F^2.
\label{f_X_Omega}
\end{equation*}
If $\mathbfcal{Y}= \mbox{\textlbrackdbl} \mathbf{U}^{(1)}, \dots, \mathbf{U}^{(N)} \mbox{\textrbrackdbl}$, then, for an arbitrary mode $i$, the corresponding matrix unfolding is given by 
	\begin{equation}
	{\bf Y}_{(i)}= \mathbf{U}^{(i)} \big(\mathbf{U}^{(N)} \odot \cdots \odot \mathbf{U}^{(i+1)}
	\odot \mathbf{U}^{(i-1)} \odot \cdots \odot \mathbf{U}^{(1)}\big)^T.
	\end{equation}
	Thus, for $i=1,\ldots,N$, $f_{\Omega}$ can be expressed as
	\begin{equation}
	f_{\Omega}(\mathbf{U}^{(1)}, \dots , \mathbf{U}^{(N)}) =\frac{1}{2}
	\,\left \| \mathbf{M}_{(i)} \circledast \left( \mathbf{X}_{(i)} - \mathbf{Y}_{(i)} \right) \right \|_F^2,
	\label{f_X_matr2}
	\end{equation}
	where $ {\bf M}_{(i)}$, and ${\bf X}_{(i)}$ are the matrix unfoldings of
	${\mathbfcal M}$ and ${\mathbfcal X}$,
	with respect to the $i$-th mode, respectively.
	These expressions form the basis of the AO algorithm for the solution of (\ref{Problem_TC}). 
Namely, we solve
\begin{equation}
\min_{{\bf U}^{(i)}\in\mathbb{R}^{I_i\times R}_+} \frac{1}{2}
\left \| \mathbf{M}_{(i)} \circledast \left( \mathbf{X}_{(i)} - \mathbf{Y}_{(i)} \right) \right \|_F^2, ~ i=1,\ldots,N. 
\label{Problem_MLS}
\end{equation}
	
	\subsection{Nonnegative Matrix Completion}
	
	We consider the NMC problem, which will be the
	building block of our AO NTC algorithm.
	Let ${\bf X}\in\mathbb{R}_+^{P \times Q}$, ${\bf A}\in\mathbb{R}_+^{P\times R}$, ${\bf B}\in\mathbb{R}_+^{Q \times R}$, $\Omega \subseteq \{1, \ldots, P \} \times \{1, \ldots, Q \} $ be the set of indices of the known entries 
	of ${\bf X}$, and ${\bf M}$ be the matrix with the same size as ${\bf X}$, with element
	${\bf M}(i,j)$ equal to one or zero based on the availability of
	the corresponding element of ${\bf X}$. We consider the problem
	\begin{equation}
	\underset{{\bf A}\in\mathbb{R}^{P\times R}_+} \min \, f_{\Omega}({\bf A}) := 
	\frac{1}{2}\,\left\| {\bf M} \circledast \left( {\bf X} - {\bf A} {\bf B}^T \right) \right \|_F^2 + \frac{\lambda}{2} \, \|{\bf A}\|_F^2.
	\label{matrix_NMC}
	\end{equation}
	The gradient and the Hessian of $f_{\Omega}$, at point ${\bf A}$, are given by
	\begin{equation}
	\nabla f_{\Omega}({\bf A}) = - \left( {\bf M} \circledast {\bf X} - {\bf M} \circledast 
	({\bf A} {\bf B}^T) \right) {\bf B} + \lambda {\bf A},
	\label{gradient_F}
	\end{equation}
	and
	\begin{equation}
	\nabla^2 f_{\Omega}({\bf A}) \hspace{-.08cm}= \hspace{-.08cm}
	( {\bf B}^T \hspace{-.08cm}\otimes \hspace{-.00cm}{\bf I}_P ) 
	\hspace{-.00cm}{\rm diag} \hspace{-.00cm}( {\rm vec} \hspace{-.00cm}( {\bf M} ) )
	( {\bf B} \otimes {\bf I}_P ) 
	\hspace{-.035cm} + \hspace{-.035cm}\lambda {\bf I}_{PR}.\hspace{-.05cm}
	\label{hessian_F}
	\end{equation}

	\subsection{Accelerated stochastic gradient for NMC}
	
	We solve problem (\ref{matrix_NMC}) via the stochastic variant of the accelerated 
	(Nesterov-type) gradient algorithm which appears in Algorithm \ref{Algorithm_StochNesterov_NMC}.
	
	During each iteration of the ``while'' loop, we use a subset of the available entries of matrix ${\bf X}$. More specifically,
	at iteration $l$, we define a set of indices 
	$\widehat{\Omega}_l \subset \Omega$ and a matrix ${\bf{\widehat{M}}}_l$, of the same size as ${\bf {M}}$, as
	\begin{equation}
	{\bf {\widehat{M}}}_l (i, j) = \left\{ \begin{array}{ll}
	1, & \mbox{if}~(i, j) \in \widehat{\Omega}_l, \\
	0, & \mbox{otherwise.} \end{array}
	\right.
	\label{M_hat_def}
	\end{equation}
	We create $\widehat{\Omega}_l$ randomly. 
	We define $B_l := |\widehat{\Omega}_l|$ and select $c:=\frac{B_l}{|\Omega|}<1$. 
	For row $p$ of matrix ${\bf X}$, for $p=1,\ldots,P$, we sample, uniformly at random, $B_{l,p}:= \floor{c  \|\mathbf{M}(p,:)\|_0}$ nonzero elements of ${\bf X}(p,:)$ ($B_{l,p}$ denotes the blocksize per row). If $B_{l,p}=0$, then we skip the $p$-th row.

	We perform an accelerated gradient step using only the elements of ${\bf X}$ whose indices appear in $\widehat{\Omega}_l$.    
	Thus, our cost function becomes $f_{\widehat{\Omega}_l}$ with gradient and Hessian similar to those in
	(\ref{gradient_F}) and (\ref{hessian_F}), with the only difference being that ${\bf M}$ is replaced by $\widehat{\bf M}_l$. 
	We find it convenient to compute the gradient and update the matrix variable in a row-wise fashion. 
	
	A novel feature of our algorithm is that each row of the matrix variable is updated via a different step-size, determined by the parameter $L_p$ (see rows $9$--$13$ of Algorithm \ref{Algorithm_StochNesterov_NMC}). 
	This can be motivated as follows.
	It is well known that the optimal step size for the gradient (or accelerated gradient) algorithm 
	for the minimization of a smooth convex function $f:\mathbb{R}^n\rightarrow \mathbb{R}$ is equal to
	$\frac{1}{L}$, where $L$ satisfies $\nabla^2 f({\bf x}) \preceq L {\bf I}$, for all ${\bf x}\in\mathbb{R}^n$.
	In line $9$ of Algorithm \ref{Algorithm_StochNesterov_NMC}, we compute ${\bf H}_{l,p}$ which is the Hessian
	of $f_{\widehat{\Omega}_l}$, with respect to the $p$-th row of ${\bf A}$. $L_p$ is the largest eigenvalue of
	${\bf H}_{l,p}$. For very sparse cases, the smallest eigenvalue of ${\bf H}_{l,p}$ is equal or very close to $\lambda$. Thus,
	the parameters used in the gradient and acceleration steps 
	(i.e., lines $11$ and $13$ of Algorithm \ref{Algorithm_StochNesterov_NMC}) 
	are those used by the constant step scheme III of
	\cite[p. 81]{Nesterov_2004}, and can be considered as ``locally optimal'' for the problem 
	at hand.

	The most demanding computations of the algorithm are as follows:
	\begin{enumerate}
		\item
		the computation of ${\bf W}_l(p,:)$ requires $O(B_{l,p} R)$ arithmetic operations (in total, $O(B_l R)$);
		
		\item
		the computation of ${\bf Z}_l(p,:)$ requires $O(B_{l,p} R)$ arithmetic operations (in total, $O(B_{l} R)$);
		
		\item
		the computation of ${\bf H}_{l,p}$ requires $O(B_{l,p} R^2)$ arithmetic operations (in total, $O(B_{l} R^2)$);
		
		\item
		the computation of $L_{p}$, via the power method, requires $O(R^2)$  arithmetic operations (in total, $O(P R^2)$).
	\end{enumerate}
	The computation of each of the matrices $\nabla f_{\widehat{\Omega}_l}$, ${\bf A}_{l+1}$ and
	${\bf Y}_{l+1}$ requires $O(PR)$ arithmetic operations.

	\begin{algorithm}[t]\small
		\DontPrintSemicolon
		\KwIn{${\bf X}, {\bf M}\hspace{-.1cm}\in\mathbb{R}_+^{P\times Q}$, ${\bf B}\hspace{-.1cm}\in\mathbb{R}_+^{Q\times R}$, ${\bf A}_*\hspace{-.1cm}\in\mathbb{R}_+^{P\times R}$, $\lambda$.}
		${\bf A}_0 = {\bf Y}_0 = {\bf A}_*$\;
		$l=0$ \;
		\While {{\rm ($l<$MAX\_INNER)}}
		{
			\For{$p = 1 \dots P $}
			{
				${\bf \widehat{M}}_l(p,:) = {\rm sample}({\bf M}(p,:))$\;
				${\bf W}_l(p,:) = -\left({\bf \widehat{M}}_l(p,:) \circledast {\bf X}(p,:) \right) {\bf B}$ \;
				${\bf Z}_l(p,:) = \left({\bf \widehat{M}}_l(p,:) \circledast \left({\bf Y}_l(p,:) {\bf B}^T\right)\right) {\bf B}$\;
				$\nabla f_{\widehat{\Omega}_l}({\bf Y}_l(p,:)) = {\bf W}_l(p,:) + {\bf Z}_l(p,:) + \lambda {\bf Y}_l(p,:)$\; 
				${\bf H}_{l, p}= {\bf B}^T {\rm diag}  \left( {\bf \widehat{M}}_l(p,:) \right){\bf B}  + \lambda {\bf I}_R$\; 		
				$L_p = {\rm max ( eig} \left( {\bf H}_{l, p} \right))$\;
				${\bf A}_{l+1}(p,:) = \left( {\bf Y}_{l}(p,:) - \frac{1}{L_p} \, \nabla f_{\widehat{\Omega}_l} ({\bf Y}_l(p,:) )\right)_+$  \;
				$\beta_{l, p} = \frac{\sqrt{L_p} - \sqrt{\lambda}}{\sqrt{L_p} + \sqrt{\lambda}}$\;
				${\bf Y}_{l+1}(p,:)  = {\bf A}_{l+1}(p,:) + \beta_{l, p} \, ({\bf A}_{l+1}(p,:)  - {\bf A}_l(p,:) )$ \;
			}
			$l=l+1$
		}
		\Return ${\bf A}_l$.
		\caption{Accelerated stochastic gradient for NMC}
		\label{Algorithm_StochNesterov_NMC}
	\end{algorithm}
	
	For notational convenience, we denote Algorithm \ref{Algorithm_StochNesterov_NMC}  as
	\begin{equation*}
	{\bf A}_{\rm opt} = {\rm S\_NMC}({\bf X}, {\bf M}, {\bf B}, {\bf A}_*, \lambda).
	\end{equation*}
We note that, in this work, ${\bf H}_{l,p}$ is used only for the determination of $L_p$. A very interesting topic is the development of algorithms that fully exploit ${\bf H}_{l,p}$. Initial efforts with a ``projected Newton step'' have not led to algorithms superior to the one presented in this paper, especially in the noisy cases. A related important topic is the development of more efficient methods for the estimation of $L_p$.
	
\section{AO accelerated stochastic NTC}
	
In order to solve the NTC problem using our accelerated stochastic algorithm, 
we start from initial values $\mathbf{U}^{(1)}_0, \ldots, {\bf U}^{(N)}_0$ and solve, 
in a circular manner, NMC problems, based on the previous estimates.
We define
	\begin{equation*}
	{\bf K}_k^{(i)} = \left(\mathbf{U}^{(N)}_k \odot \dots \odot \mathbf{U}^{(i+1)}_k \odot \mathbf{U}^{(i-1)}_{k+1} \odot \dots \odot \mathbf{U}^{(1)}_{k+1} \right),
	\end{equation*}
	where $k$ denotes the $k$--th AO iteration.
	The update of ${\bf U}^{(i)}_k$ is attained by the function call ${\rm S\_NMC}({\bf X}_{(i)}, {\bf M}_{(i)}, {\bf K}_k^{(i)}, {\bf U}^{(i)}_k, \lambda)$. The Stochastic NTC algorithm appears in Algorithm \ref{Algorithm_StochasticNALS_Completion}. 
	
	\begin{algorithm}[t]
		\DontPrintSemicolon
		\KwIn{${\mathbfcal X}$, $\Omega$, $ \left\lbrace \mathbf{U}^{(i)}_0 \right\rbrace_{i = 1}^{N}$, $\lambda$, \textit{R}.}
		$k=0$\;
		\While{$(1)$} {
			\For{$i = 1,2, \dots N$}{
				$\mathbf{U}^{(i)}_{k+1}  = {\rm S\_NMC}\left({\bf X}_{(i)}, {\bf M}_{(i)},{\bf K}_{k}^{(i)}, {\bf U}^{(i)}_k, \lambda \right)$  \;
			}
			{\bf if~}(term\_cond is TRUE) {\bf then} {\rm break;} {\bf endif}\;
			$k=k+1$\;}
		\Return $\left\lbrace \mathbf{U}^{(i)}_k \right\rbrace_{i = 1}^{N}$.
		\caption{AO accelerated stochastic NTC}
		\label{Algorithm_StochasticNALS_Completion}
	\end{algorithm}

	\subsection{Parallel Implementation of AO accelerated stochastic NMC}
	
	In Algorithm \ref{Algorithm_ParallelStochNesterov_NMC}, we provide a high level algorithmic sketch of the
	accelerated stochastic gradient for NMC. We employ the OpenMP API, which is suitable for our multi-threading approach. The update of each row can be done independently. Therefore, lines $5$-$13$ of Algorithm \ref{Algorithm_StochNesterov_NMC}, can be computed separately by each available thread.
	
	\begin{algorithm}[t]
		\DontPrintSemicolon
		\KwIn{${\bf X}, {\bf M}\hspace{-.1cm}\in\mathbb{R}_+^{P\times Q}$, ${\bf B}\hspace{-.1cm}\in\mathbb{R}_+^{Q\times R}$, ${\bf A}_*\hspace{-.1cm}\in\mathbb{R}_+^{P\times R}$, $\lambda$}	${\bf A}_0 = {\bf Y}_0 = {\bf A}_*$\;
		$l=0$ \;
		\While {$(1)$}{
			\uIf{{\rm ($l\geq$MAX\_INNER)}}{
				{\rm break} \;
			}
			\Else{
				\textbf{in parallel}
				\For{$p = 1 \dots P $}{
					lines 5-13 of Algorithm \ref{Algorithm_StochNesterov_NMC} \;
				}
				$l=l+1$
			}
		}
		\Return ${\bf A}_l$.
		\caption{Parallel accelerated stochastic gradient for NMC}
		\label{Algorithm_ParallelStochNesterov_NMC}
	\end{algorithm}

	\section{Numerical Experiments}
	
	In this section, we test the effectiveness of our algorithm in various test cases, using both real-world and synthetic data. 
	We denote as epoch the number of iterations required to access once all available tensor elements. In 
	subsections \ref{subsection_convergence_speed_stochastic_NTC} and \ref{subsection_execution_time_parallel}, we 
	compute averages over $5$ Monte Carlo trials. 
	
	\subsection{Stochastic NTC on corrupted image }
	
\begin{figure}[t]
		\begin{subfigure}{0.5\textwidth}\centering
			\includegraphics[scale=0.1]{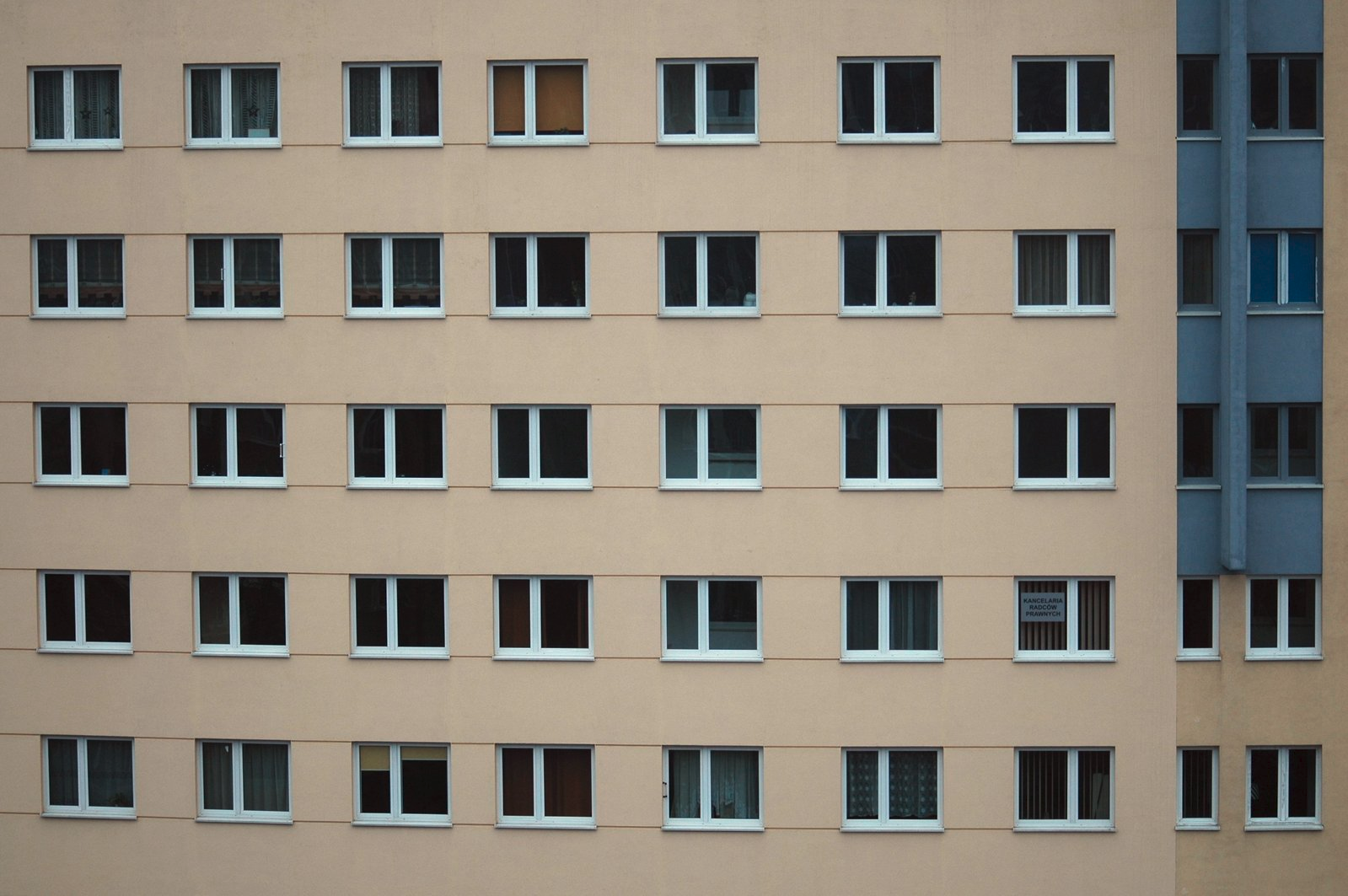}
			\caption{Original}
			\label{fig:original_img} 
		\end{subfigure}
		\begin{subfigure}{0.5\textwidth}\centering
			\includegraphics[scale=0.1]{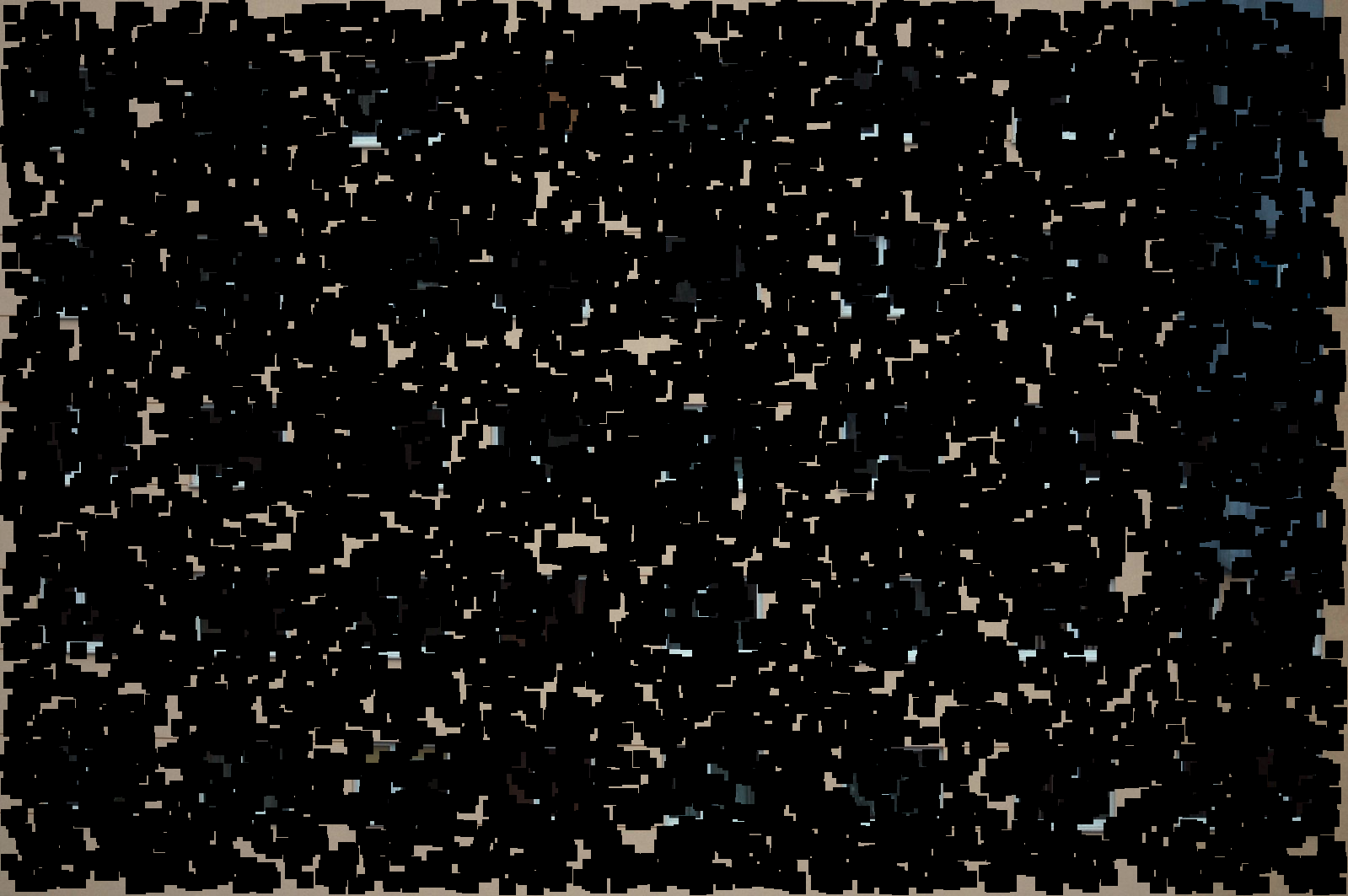}
			\caption{Corrupted ($90\%$ sparse)}
			\label{fig:damaged_img} 
		\end{subfigure}
		\begin{subfigure}{0.5\textwidth}\centering
			\includegraphics[scale=0.1]{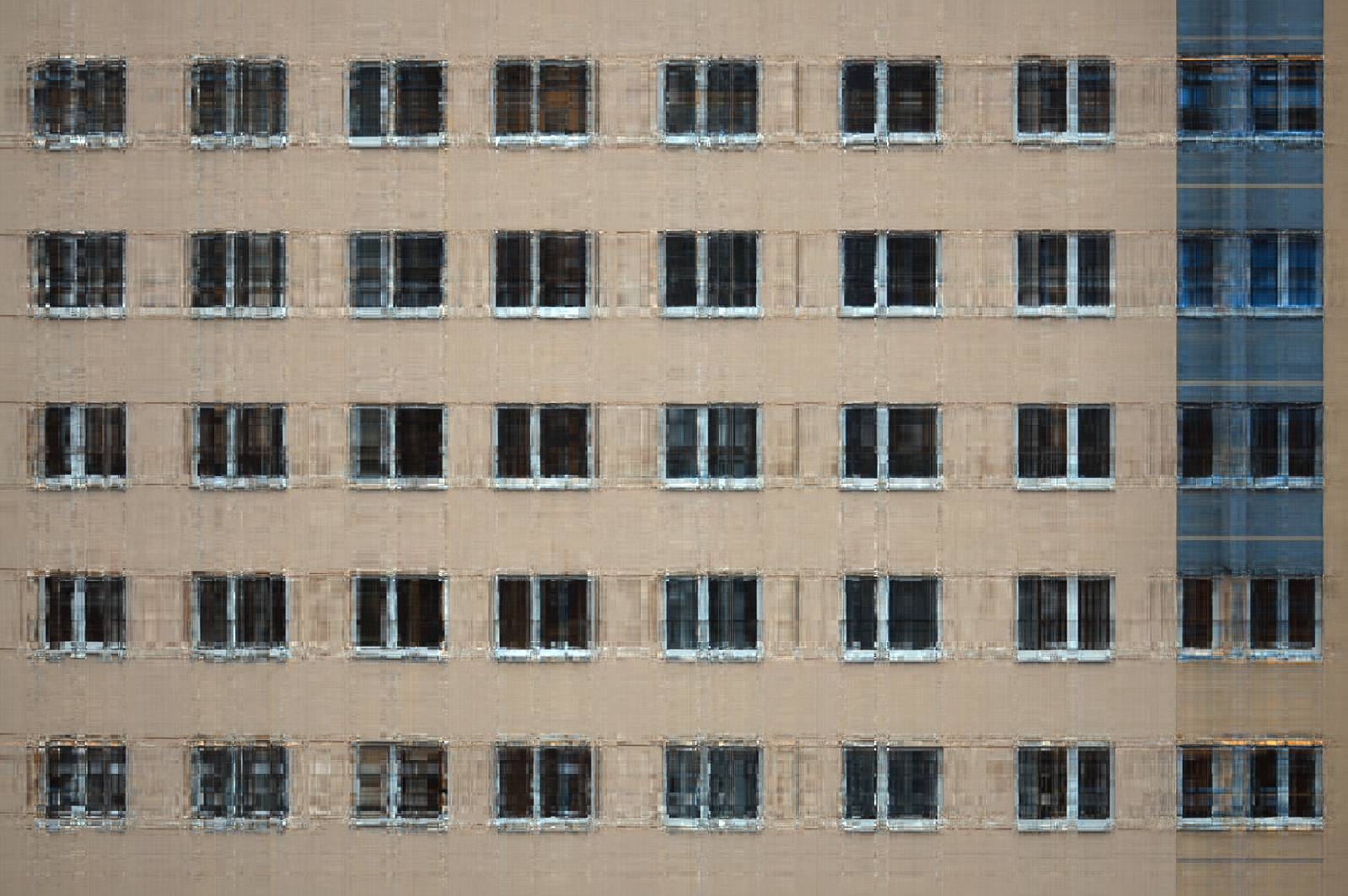}
			\caption{Restored}
			\label{fig:restored_img}
		\end{subfigure}
		\caption{Tensor Completion on a corrupted image.}
		\label{fig:image_completion}
	\end{figure}

	We start by estimating missing data in images,\footnote{Image can be found at https://images.freeimages.com/images/large-previews/7bb/building-1222550.jpg} 
	following the RGB model. In Fig. \ref{fig:image_completion}, we depict the results obtained from the application of the AO Stochastic NTC on a corrupted image of dimensions $1063 \times 1599 \times 3$ ($90\%$ sparse). We set $c=0.02$, number of epochs $500$, ${\rm MAX\_INNER}=1$, and rank $R=50$. We observe that the algorithm is able to reconstruct the image even for small values of $c$.
	
	\subsection{Convergence speed of Stochastic NTC}
	\label{subsection_convergence_speed_stochastic_NTC}
	
\begin{figure}[t]
		\centering
		\includegraphics[scale=0.5]{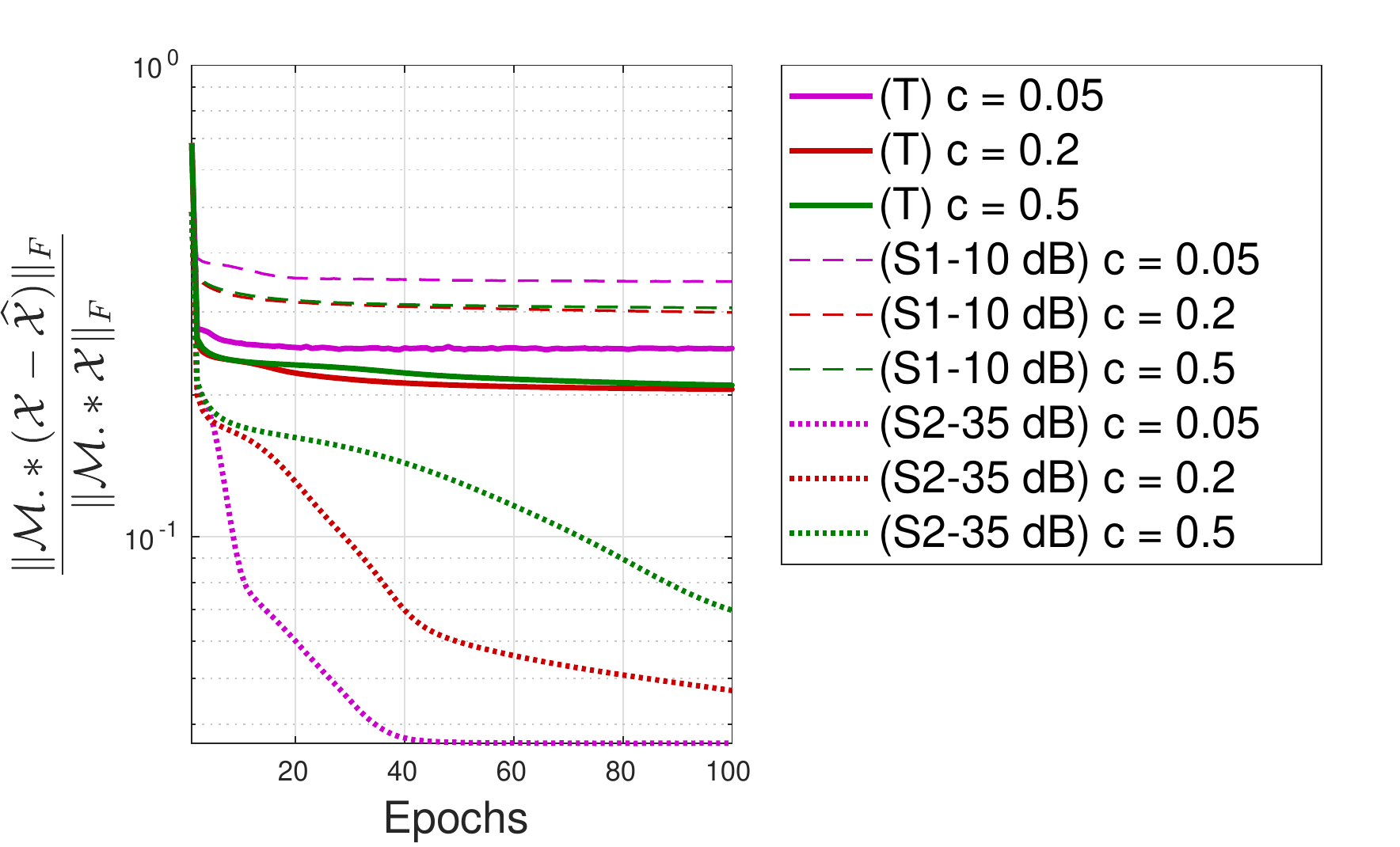}
		\caption{Relative Reconstruction Error vs Epochs for real-world (T) and noisy synthetic (S1 and S2) data, using various values of $c=0.05, 0.2, 0.5$.}
		\label{fig:convergence}
	\end{figure}

	We experimentally evaluate the convergence speed of our algorithm using both real-world and synthetic data. Concerning the real-world data, we use the dataset 
	``MovieLens 10M'' \cite{Movielens-Harper_2015} of dimensions $71567 \times 65133 \times 730$.
	Concerning the synthetic data, we generate a rank-$10$ nonnegative
	tensor $\mathbfcal{X}^o$, of size equal to the real-world data, whose true latent factors have independent and identically distributed (i.i.d.) elements, drawn from $\mathcal{U}[0, 1]$. The additive noise $\mathbfcal{E}$ has i.i.d. elements $\mathbfcal{N}(0,\sigma^2_N)$. In order to create a tensor of the same sparsity level as the real-world dataset, we generate a tensor $\mathbfcal{M}$ of the same size as $\mathbfcal{X}^o$ according to (\ref{M_def}). The observed incomplete tensor is expressed as $\mathbfcal{M} \circledast \left(\mathbfcal{X}^o + \mathbfcal{E}\right)$. We define the
	Signal-to-Noise ratio as
	\begin{equation}
	{\rm SNR}:=\frac{\|  \mathbfcal{M}  \circledast \mathbfcal{X}^o\|_F^2}
	{\| \mathbfcal{M} \circledast  \mathbfcal{E} \|_F^2}.
	\end{equation}
	In Fig. \ref{fig:convergence}, we plot the average Relative Reconstruction Error for $100$ epochs, ${\rm MAX\_INNER}=1$ and rank $R=10$. We observe that, in noisy environments, choosing small values of $c$ is not effective. On the contrary, in the high SNR cases, small values of 
	$c$ are more suitable.
	
	\begin{figure}[t]
		\centering
		\includegraphics[scale=0.5]{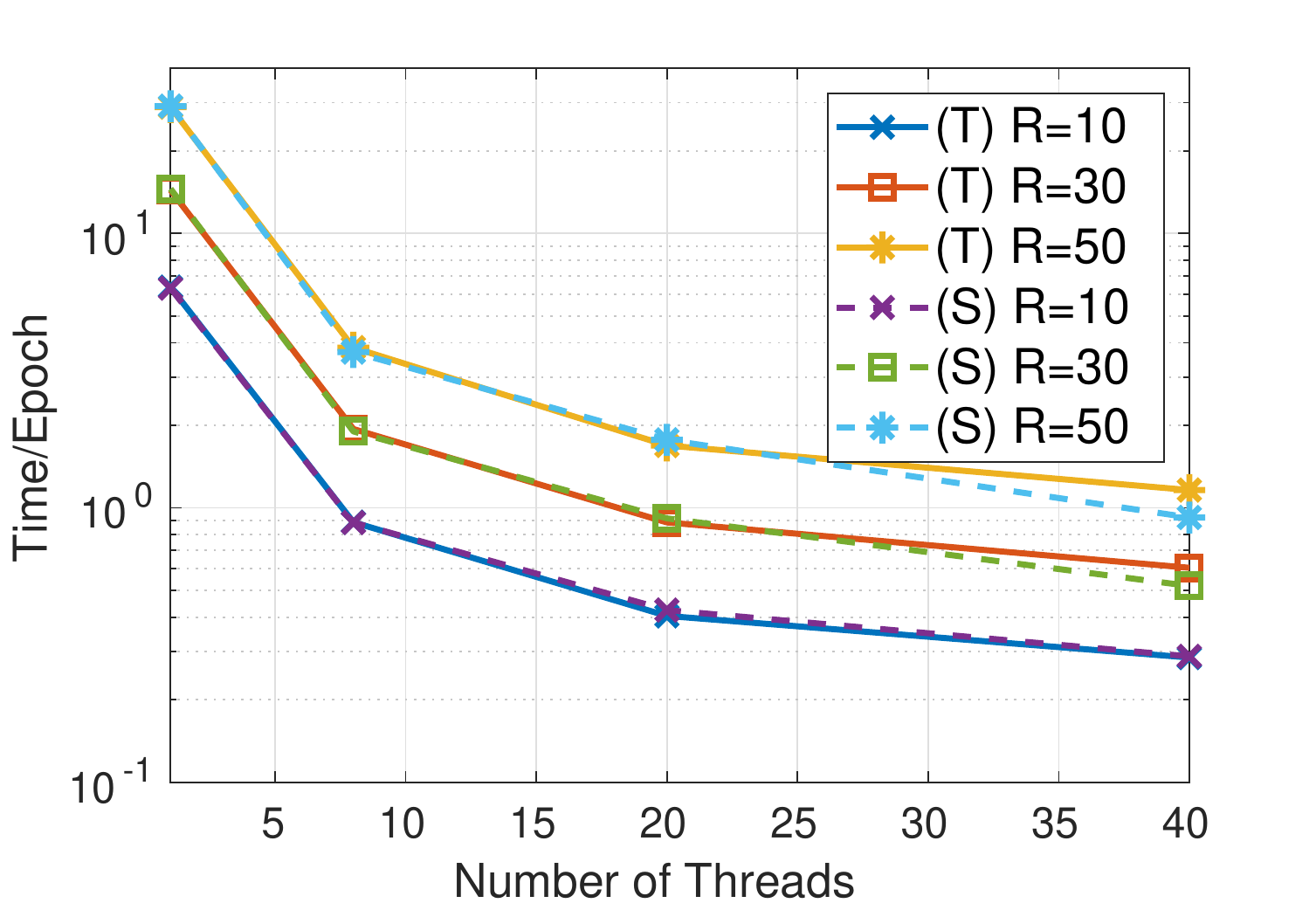}
		\caption{Execution time per epoch (in seconds) for synthetic (S) and real-world (T) dataset with dimensions $183 \times 24 \times 1140 \times 1717$ and $c=0.5$, for various values of rank $R$, versus number of threads.}
		\label{fig:execution_time}
	\end{figure}
	
	\begin{figure}[t]
		\centering
		\includegraphics[scale=0.5]{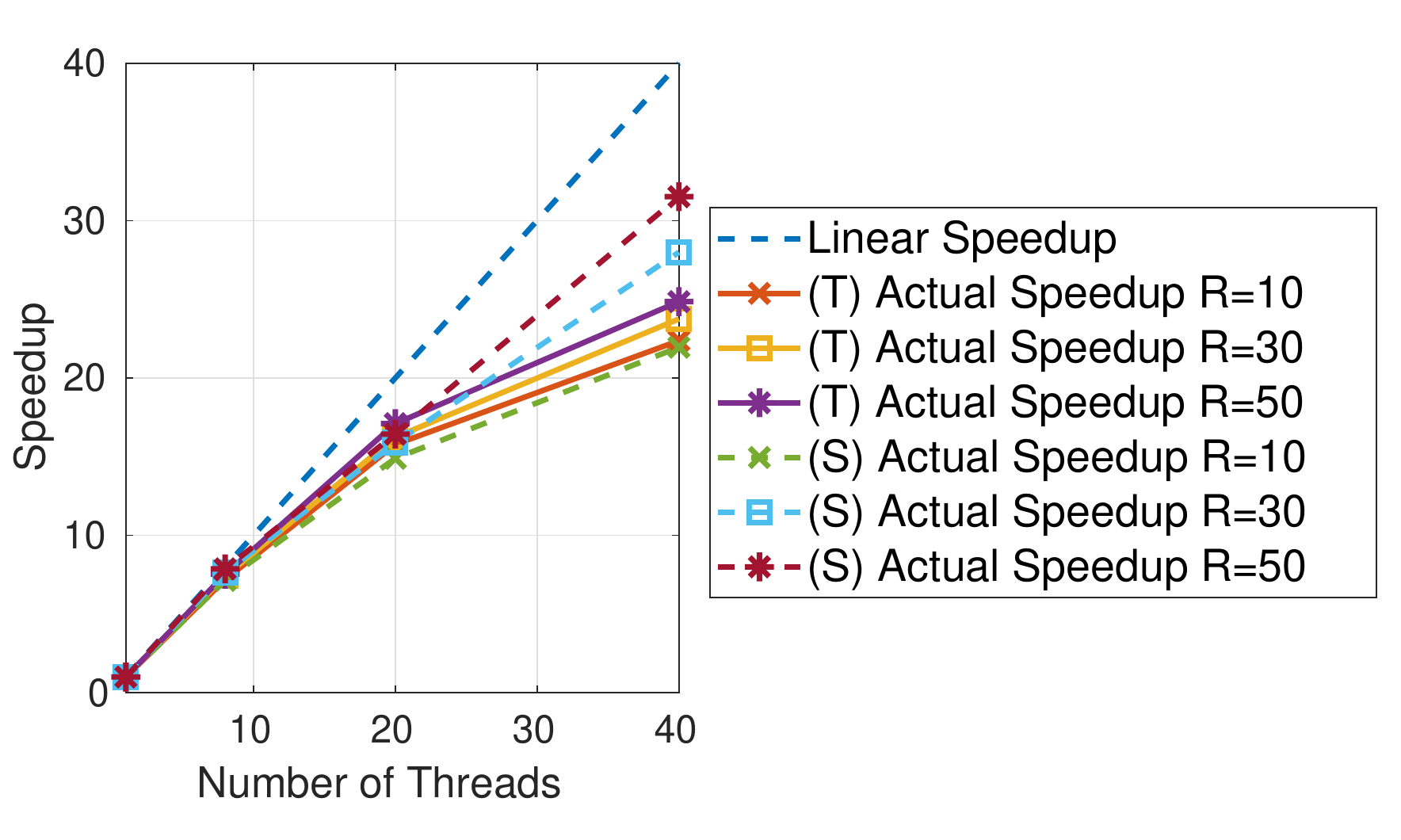}
		\caption{Speedup attained for synthetic (S) and real-world (T) dataset with dimensions $183 \times 24 \times 1140 \times 1717$ and $c=0.5$, for various values of rank $R$, versus number of threads.}
		\label{fig:speedup}
	\end{figure}
	
	\subsection{Execution time for parallel stochastic NTC}
	\label{subsection_execution_time_parallel}
	
	We assess the performance of our algorithm in a shared-memory environment using 
	both real-world and synthetic data (of the same dimensions). Concerning the real-world data,
	we use the dataset ``Uber Pickups'' \cite{frosttdataset}, which can be represented as a $4$--th order tensor of dimensions $183 \times 24 \times 1140 \times 1717$ and $3309490$ nonzero elements. For the synthetic data, we create a
	rank-$10$ tensor $\mathbfcal{X}^o$, whose latent factors have i.i.d elements ${\cal U}[0,1]$. Similarly, we generate a tensor
	$\mathbfcal{M}$ of the same size with $\mathbfcal{X}^o$. Thus, the observed incomplete and noiseless tensor is $\mathbfcal{M} \circledast \mathbfcal{X}^o$ with the same number of nonzeros as in the ``Uber Pickups'' dataset. We set $c=0.5$, number of epochs $1$, ${\rm MAX\_INNER}=1$. In order to test the algorithm's performance, we examine various values of the rank $R = 10,30,50$. In Fig. \ref{fig:execution_time}, we illustrate the average execution time per epoch versus the number of threads. In Fig. \ref{fig:speedup}, we present the average attained speedup. We observe that significant speedup is attained in all cases. We also observe that the speedup attained with synthetic data is somewhat higher than that attained with real-world data. This happens due to load imbalancing, caused by the nonuniform distribution of the nonzero elements of the real-world data, among the available threads.
	
	\section{Conclusion}
	
	We considered the NTC problem. First, we developed an accelerated stochastic algorithm for the NMC problem.
	A unique feature of our approach is that each row of the matrix variable is updated using a different step-size, 
	specifically tailored to this row. Then, we used this algorithm and built an AO algorithm for the NTC problem. 
	We tested the data reconstruction effectiveness as well as the convergence speed of our approach using both synthetic and real-world data. We implemented our algorithm using the OpenMP API, and observed significant speedup. Our method
	is an effective and efficient candidate for the solution of very large-scale NTC problems.

	\section*{Acknowledgment}
	This work was also supported by computational time granted from the National Infrastructures for Research and Technology S.A. (GRNET S.A.) in the National HPC facility - ARIS - under project ID PR008040–PARTENSOR SPARSE.
	
	%
	%
	%
	
	\bibliographystyle{IEEEtran}

\begin{thebibliography}{10}
\providecommand{\url}[1]{#1}
\csname url@samestyle\endcsname
\providecommand{\newblock}{\relax}
\providecommand{\bibinfo}[2]{#2}
\providecommand{\BIBentrySTDinterwordspacing}{\spaceskip=0pt\relax}
\providecommand{\BIBentryALTinterwordstretchfactor}{4}
\providecommand{\BIBentryALTinterwordspacing}{\spaceskip=\fontdimen2\font plus
\BIBentryALTinterwordstretchfactor\fontdimen3\font minus
  \fontdimen4\font\relax}
\providecommand{\BIBforeignlanguage}[2]{{%
\expandafter\ifx\csname l@#1\endcsname\relax
\typeout{** WARNING: IEEEtran.bst: No hyphenation pattern has been}%
\typeout{** loaded for the language `#1'. Using the pattern for}%
\typeout{** the default language instead.}%
\else
\language=\csname l@#1\endcsname
\fi
#2}}
\providecommand{\BIBdecl}{\relax}
\BIBdecl

\bibitem{Kroonenberg_2008}
P.~M. Kroonenberg, \emph{Applied Multiway Data Analysis}.\hskip 1em plus 0.5em
  minus 0.4em\relax Wiley-Interscience, 2008.

\bibitem{Cichocki_et_al_2009}
A.~Cichocki, R.~Zdunek, A.~H. Phan, and S.~Amari, \emph{Nonnegative Matrix and
  Tensor Factorizations}.\hskip 1em plus 0.5em minus 0.4em\relax Wiley, 2009.

\bibitem{Kolda_Bader_2009}
T.~G. Kolda and B.~W. Bader, ``Tensor decompositions and applications,''
  \emph{SIAM Review}, vol.~51, no.~3, pp. 455--500, September 2009.

\bibitem{Sidiropoulos_et_al_2017}
N.~D. Sidiropoulos, L.~De~Lathauwer, X.~Fu, K.~Huang, E.~E. Papalexakis, and
  C.~Faloutsos, ``Tensor decomposition for signal processing and machine
  learning,'' \emph{IEEE Transactions on Signal Processing}, vol.~65, no.~13,
  pp. 3551--3582, 2017.

\bibitem{Nesterov_2004}
Y.~Nesterov, \emph{Introductory lectures on convex optimization}.\hskip 1em
  plus 0.5em minus 0.4em\relax Kluwer Academic Publishers, 2004.

\bibitem{Kang_Papalexakis_et_al_2012}
U.~Kang, E.~Papalexakis, A.~Harpale, and C.~Faloutsos, ``Gigatensor: Scaling
  tensor analysis up by 100 times - algorithms and discoveries,''
  \emph{Proceedings of the 18th ACM SIGKDD International Conference on
  Knowledge Discovery and Data Mining (KDD12) Beiging, China}, 2012.

\bibitem{Choi_Vishwanathan_2014}
J.~H. Choi and S.~V.~N. Vishwanathan, ``Dfacto: Distributed factorization of
  tensors,'' \emph{Advances in Neural Information Processing Systems (NIPS)},
  2014.

\bibitem{Karlsson_et_al_2015}
L.~Karlsson, D.~Kressner, and A.~Uschmajew, ``Parallel algorithms for tensor
  completion in the {CP} format,'' \emph{Parallel Computing}, 2015.

\bibitem{Smith_Karypis_2015}
S.~Smith and G.~Karypis, ``A medium-grained algorithm for distributed sparse
  tensor factorization,'' \emph{30th IEEE International Parallel \& Distributed
  Processing Symposium}, 2016.

\bibitem{Kaya_et_al_2018}
O.~Kaya and B.~U{\c{c}}ar, ``Parallel candecomp/parafac decomposition of sparse
  tensors using dimension trees,'' \emph{SIAM Journal on Scientific Computing},
  vol.~40, no.~1, pp. C99--C130, 2018.

\bibitem{Huang_Sidiropoulos_Liavas_2015}
K.~Huang, N.~D. Sidiropoulos, and A.~P. Liavas, ``A flexible and efficient
  framework for constrained matrix and tensor factorization,'' \emph{IEEE
  Transactions on Signal Processing}, accepted for publication, May 2016.

\bibitem{smith_karypis_AO_ADMM2015}
S.~Smith, A.~Beri, and G.~Karypis, ``Constrained tensor factorization with
  accelerated ao-admm,'' \emph{2017 45th International Conference on Parallel
  Processing (ICPP), Bristol}, 2017.

\bibitem{Liavas_et_al_2018}
A.~P. Liavas, G.~Kostoulas, G.~Lourakis, K.~Huang, and N.~D. Sidiropoulos,
  ``Nesterov-based alternating optimization for nonnegative tensor
  factorization: Algorithm and parallel implementations,'' \emph{IEEE
  Transactions on Signal Processing}, vol.~66, no.~4, pp. 944--953, Feb. 2018.

\bibitem{Ballard_et_al_2018}
G.~Ballard, K.~Hoyashi, and R.~Kannan, ``Parallel nonnegative cp decompositions
  of dense tensors,'' \emph{2018 IEEE 25th International Conference on High
  Prformance Computing (HiPC), Bengaluru, India}, 2018.

\bibitem{Blanco_et_al_2018}
\BIBentryALTinterwordspacing
Z.~Blanco, B.~Liu, and M.~M. Dehnavi, ``Cstf: Large-scale sparse tensor
  factorizations on distributed platforms,'' in \emph{Proceedings of the 47th
  International Conference on Parallel Processing}, ser. ICPP 2018.\hskip 1em
  plus 0.5em minus 0.4em\relax New York, NY, USA: Association for Computing
  Machinery, 2018. [Online]. Available:
  \url{https://doi.org/10.1145/3225058.3225133}
\BIBentrySTDinterwordspacing

\bibitem{Ge_et_al_2018}
H.~{Ge}, K.~{Zhang}, M.~{Alfifi}, X.~{Hu}, and J.~{Caverlee}, ``Distenc: A
  distributed algorithm for scalable tensor completion on spark,'' in
  \emph{2018 IEEE 34th International Conference on Data Engineering (ICDE)},
  2018, pp. 137--148.

\bibitem{Shin_Kang_2014}
K.~Shin and U.~Kang, ``Distributed methods for high-dimensional and large-scale
  tensor factorization,'' in \emph{{IEEE} International Conference on Data
  Mining, {ICDM}}, 2014, pp. 989--994.

\bibitem{Karsavuran_et_al_2021}
M.~O. {Karsavuran}, S.~{Acer}, and C.~{Aykanat}, ``Partitioning models for
  general medium-grain parallel sparse tensor decomposition,'' \emph{IEEE
  Transactions on Parallel and Distributed Systems}, vol.~32, no.~1, pp.
  147--159, 2021.

\bibitem{Papastergiou_Megalooikonomou_2017}
T.~{Papastergiou} and V.~{Megalooikonomou}, ``A distributed proximal gradient
  descent method for tensor completion,'' in \emph{2017 IEEE International
  Conference on Big Data (Big Data)}, 2017, pp. 2056--2065.

\bibitem{Smith_et_al_2016}
S.~{Smith}, J.~{Park}, and G.~{Karypis}, ``An exploration of optimization
  algorithms for high performance tensor completion,'' in \emph{SC '16:
  Proceedings of the International Conference for High Performance Computing,
  Networking, Storage and Analysis}, 2016, pp. 359--371.

\bibitem{Xie_et_al_2020}
K.~Xie, Y.~Chen, G.~Wang, G.~Xie, J.~Cao, and J.~Wen, ``Accurate and fast
  recovery of network monitoring data: A gpu accelerated matrix completion,''
  \emph{IEEE/ACM Transactions on Networking}, vol.~PP, pp. 1--14, 03 2020.

\bibitem{Devine_Ballard_2020}
K.~Devine and G.~Ballard, ``{GentenMPI}: Distributed memory sparse tensor
  decomposition,'' Tech. Rep. SAND2020-8515, 2020.

\bibitem{Movielens-Harper_2015}
\BIBentryALTinterwordspacing
F.~M. Harper and J.~A. Konstan, ``The movielens datasets: History and
  context,'' \emph{ACM Trans. Interact. Intell. Syst.}, vol.~5, no.~4, Dec.
  2015. [Online]. Available: \url{https://doi.org/10.1145/2827872}
\BIBentrySTDinterwordspacing

\bibitem{frosttdataset}
\BIBentryALTinterwordspacing
S.~Smith, J.~W. Choi, J.~Li, R.~Vuduc, J.~Park, X.~Liu, and G.~Karypis. (2017)
  {FROSTT}: The formidable repository of open sparse tensors and tools.
  [Online]. Available: \url{http://frostt.io/}
\BIBentrySTDinterwordspacing

\end{thebibliography}

	\vspace{12pt}
	
\end{document}